\documentclass[aip, amsmath,amssymb,reprint,floatfix]{revtex4-2}
\usepackage[utf8]{inputenc}
\usepackage[T1]{fontenc}
\usepackage{graphicx}
\usepackage{dcolumn, lipsum,graphicx,float}
\usepackage{diagbox}
\graphicspath{ {./Figures/} }
\usepackage{slashbox}
\usepackage{gensymb}
\usepackage{wrapfig}
\usepackage[dvipsnames]{xcolor}
\usepackage{hyperref}
\usepackage[modulo,pagewise]{lineno}
\usepackage{caption}
\usepackage{subcaption}
\usepackage{verbatim}
\usepackage[newcommands]{ragged2e}
\usepackage[export]{adjustbox}

\hypersetup{
    colorlinks=true,
    linkcolor=blue,
    citecolor = blue,
    urlcolor=blue
}

\usepackage{color}
\usepackage{soul}
\newcommand{\Eq}[1]{Eq.~\eqref{#1}}
\newcommand{\no}{\nonumber\\}

\begin{document}

\title{Three-wave mixing traveling-wave parametric amplifier with periodic variation of the circuit parameters}

\author{Anita Fadavi Roudsari}
\email{fadavi@chalmers.se}
\author{Daryoush Shiri, Hampus Renberg Nilsson, Giovanna Tancredi, Amr Osman}
\author{Ida-Maria Svensson, Marina Kudra, Marcus Rommel, Jonas Bylander, Vitaly Shumeiko}
\author{Per Delsing}

\affiliation{Department of Microtechnology and Nanoscience, Chalmers University of Technology, 412 96 Gothenburg, Sweden}

\date{\today}

\begin{abstract}
We report the implementation of a near-quantum-limited, traveling-wave parametric amplifier that uses three-wave mixing (3WM). 
To favor amplification by 3WM, we use the superconducting nonlinear asymmetric inductive element (SNAIL) loops, biased with a dc magnetic flux.
In addition, we equip the device with dispersion engineering features to create a stop-band at the second harmonic of the pump and suppress the propagation of the higher harmonics that otherwise degrade the amplification. With a chain of 440 SNAILs, the amplifier provides up to $20\,\text{dB}$ gain and a 3-dB bandwidth of $1\,\text{GHz}$. The added noise by the amplifier is found to be less than one photon.  

\end{abstract}

\maketitle
Over six decades have passed since the first proposals on traveling-wave parametric amplifiers (TWPAs). \cite{Cullen_1958,tien_1958,tien_suhl_1958} Since then, TWPAs have been widely used in optics;  \cite{Hedekvist_1997,Kylemark_2006} however, in electronics and microwave circuits, they gave way to amplifiers made of transistors. Recent progress in superconducting quantum computing and the demand for quantum-limited amplifiers has brought the attention back to the TWPAs again. Being composed of passive and low-loss elements such as capacitors and inductors, these amplifiers are inherently low noise, and this makes them ideal for millikelvin measurements. Furthermore, embedding the elements in a transmission-line frame liberates TWPAs from the constraint of a fixed gain-bandwidth product inherent in parametric amplifiers based on lumped element oscillators and resonators. \cite{yamamoto_2008,sundqvist_2013}

Parametric amplification results from frequency mixing in a nonlinear element, where a strong pump is mixed with the weak signal, causing energy transfer from the pump to the signal and hence amplification. There are two main schemes of frequency mixing, namely three-wave mixing (3WM) and four-wave mixing (4WM). During 3WM, one photon at the pump frequency $\omega_p$ generates a photon at the signal frequency 
$\omega_s$ and another photon at an idler frequency $\omega_i$ under the resonance condition $\omega_p = \omega_s + \omega_i$, whereas during 4WM, two photons at the pump frequency create the signal and the idler photons, i.e., $2\omega_p = \omega_s + \omega_i$.
Among the two schemes, 4WM-TWPAs appeared earlier and progressed both in the form of kinetic-inductance-based transmission lines \cite{eom_2012,bockstiegel_2014,adamyan_2016,chaudhuri_2017,goldstein_2020} and of lumped-element circuits composed of Josephson junctions for the required nonlinearity. \cite{TWPA_mohebbi,Yaakobi_2013, macklin_2015, white_2015,Planat_2020,ranadive_2022} There has been fewer implementations of 3WM-TWPAs, \cite{zorin_2017, vissers_2016,ranzani_2018,Miano_2019,shu_2021, malnou_2021, perelshtein_2021} despite the abundance of the theoretical work. \cite{tien_1958,zorin_2016,Erickson_2017,zorin_2017,zorin_2019,Dixon_2020,zorin_2021}

The quadratic nonlinearity in a 3WM-TWPA leads to a greater amplification at lower pump powers compared with a 4WM-TWPA which relies on cubic nonlinearity. This translates into a shorter chain for the 3WM-TWPA to achieve a target amplification, \cite{zorin_2016} which in turn introduces less loss to the system upon implementation. In addition, in a 3WM-TWPA, maximum amplification takes place close to half of the pump frequency, in contrast to a 4WM-TWPA, where the gain is maximum near the pump frequency. Therefore, the strong pump that is detuned from the signal can be filtered out easily. Furthermore, the 3WM-TWPA does not suffer from the intrinsic phase mismatch induced by the Kerr effect, which was believed to eliminate the need for dispersion engineering. \cite{zorin_2016,zorin_2017}
In practice, however, the linear dispersion also facilitates the generation of higher pump harmonics and up-converted signal modes, and therefore, the desired gain may not be achieved unless one increases the number of unit cells in the device. \cite{Dixon_2020,nilsson_2022}
This problem has so far been addressed by adding features to the TWPA chain to deliberately distort the linear dispersion and increase the phase mismatch at higher frequencies while maintaining finely tuned phase matching for a particular pump frequency. Although this approach results in gain enhancement, it suffers from the limited flexibility of the amplification band. \cite{vissers_2016,malnou_2021} 
\begin{figure*}
     \centering
     \captionsetup{justification=Justified}
     \begin{subfigure}[b]{0.22\textwidth}
         \centering
         \includegraphics[width=\textwidth]{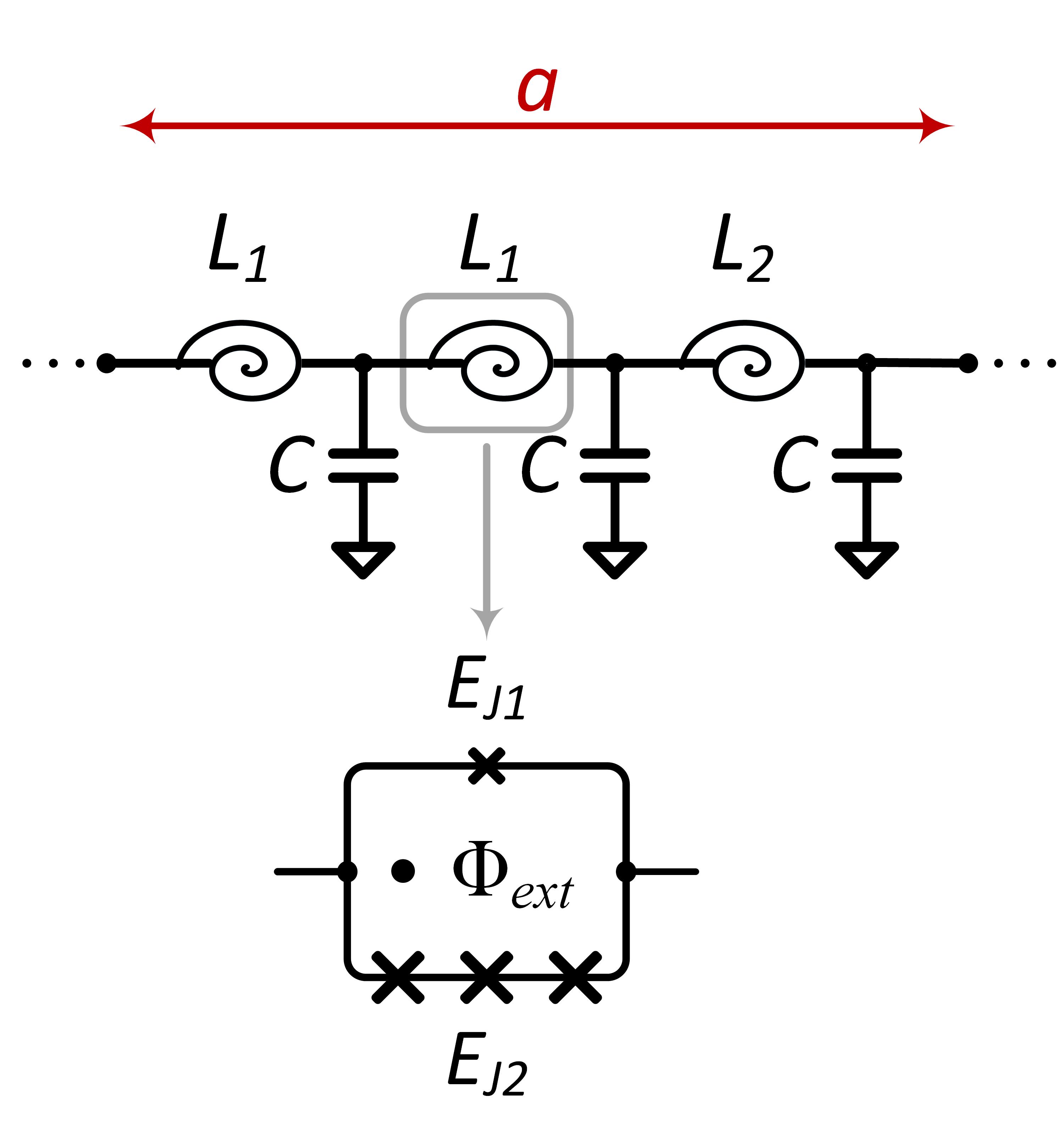}
         \caption{}
         \label{fig:TWPA_UC}
     \end{subfigure}
     \hfill
     \begin{subfigure}[b]{0.22\textwidth}
         \centering
         \includegraphics[width=\textwidth]{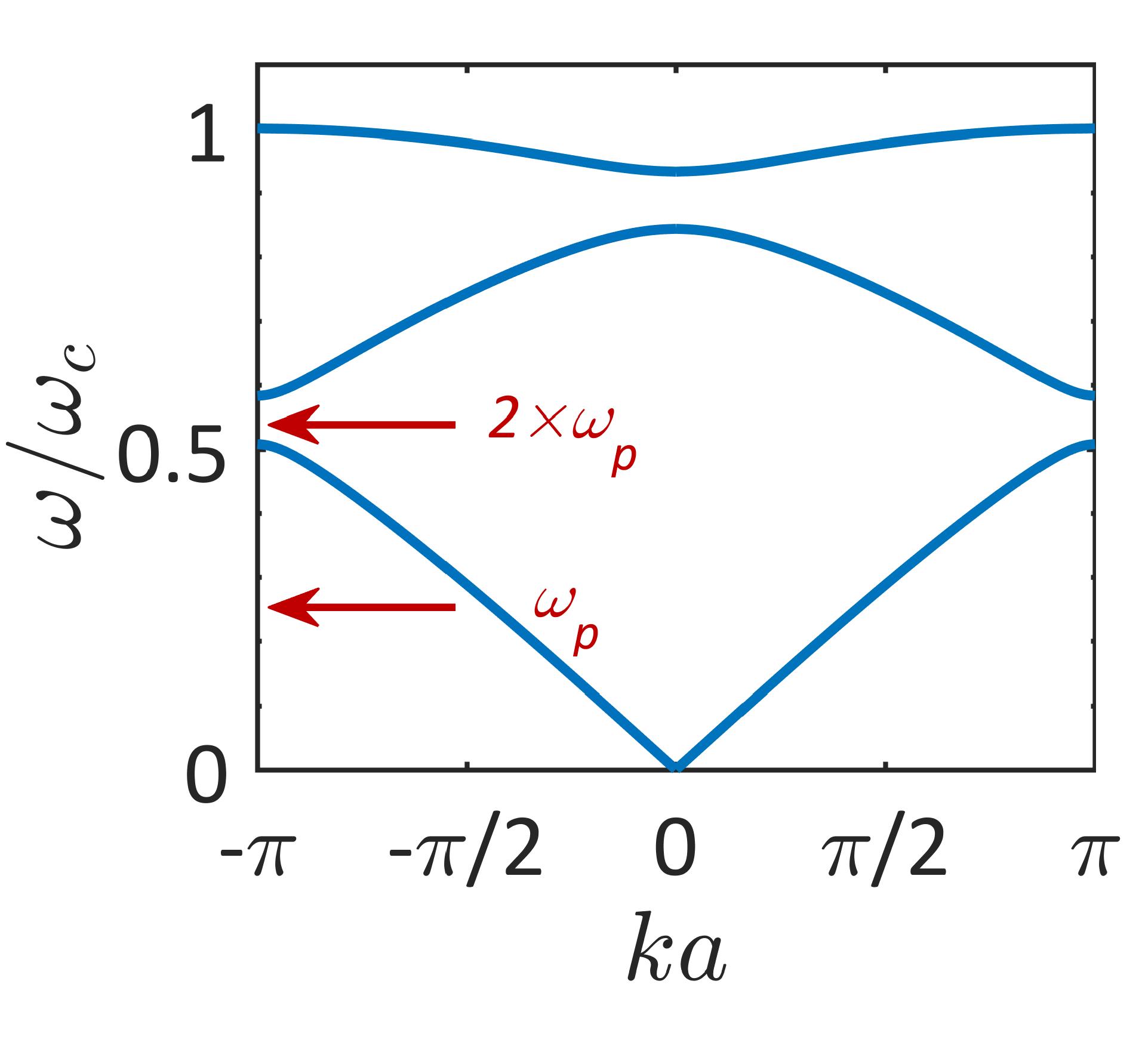}
         \caption{}
         \label{fig:dispersion}
     \end{subfigure}
     \hfill
     \begin{subfigure}[b]{0.27\textwidth}
         \centering
         \includegraphics[width=\textwidth]{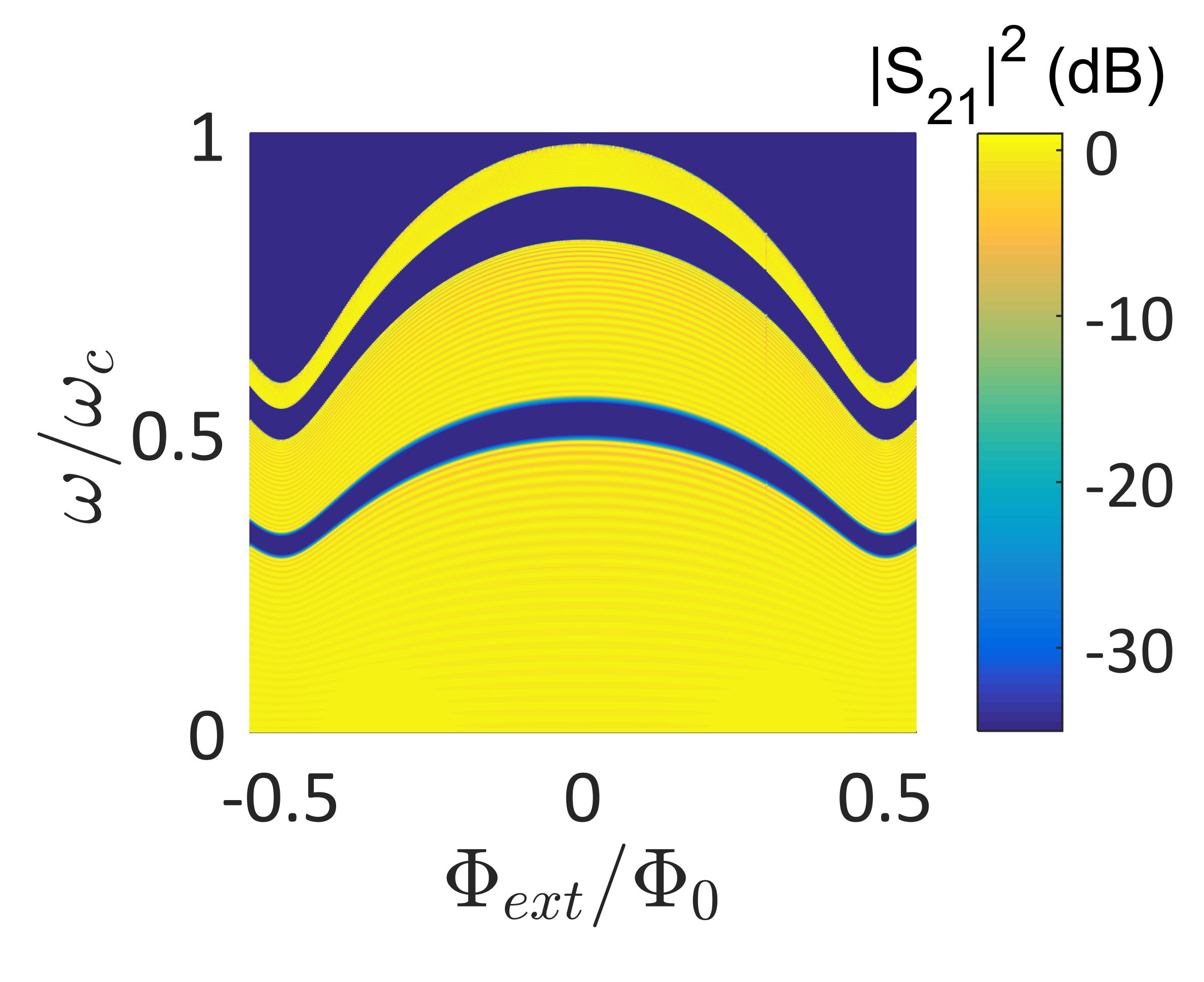}
         \caption{}
         \label{fig:S21}
     \end{subfigure}
     \hfill
     \begin{subfigure}[b]{0.22\textwidth}
         \centering
         \includegraphics[width=\textwidth]{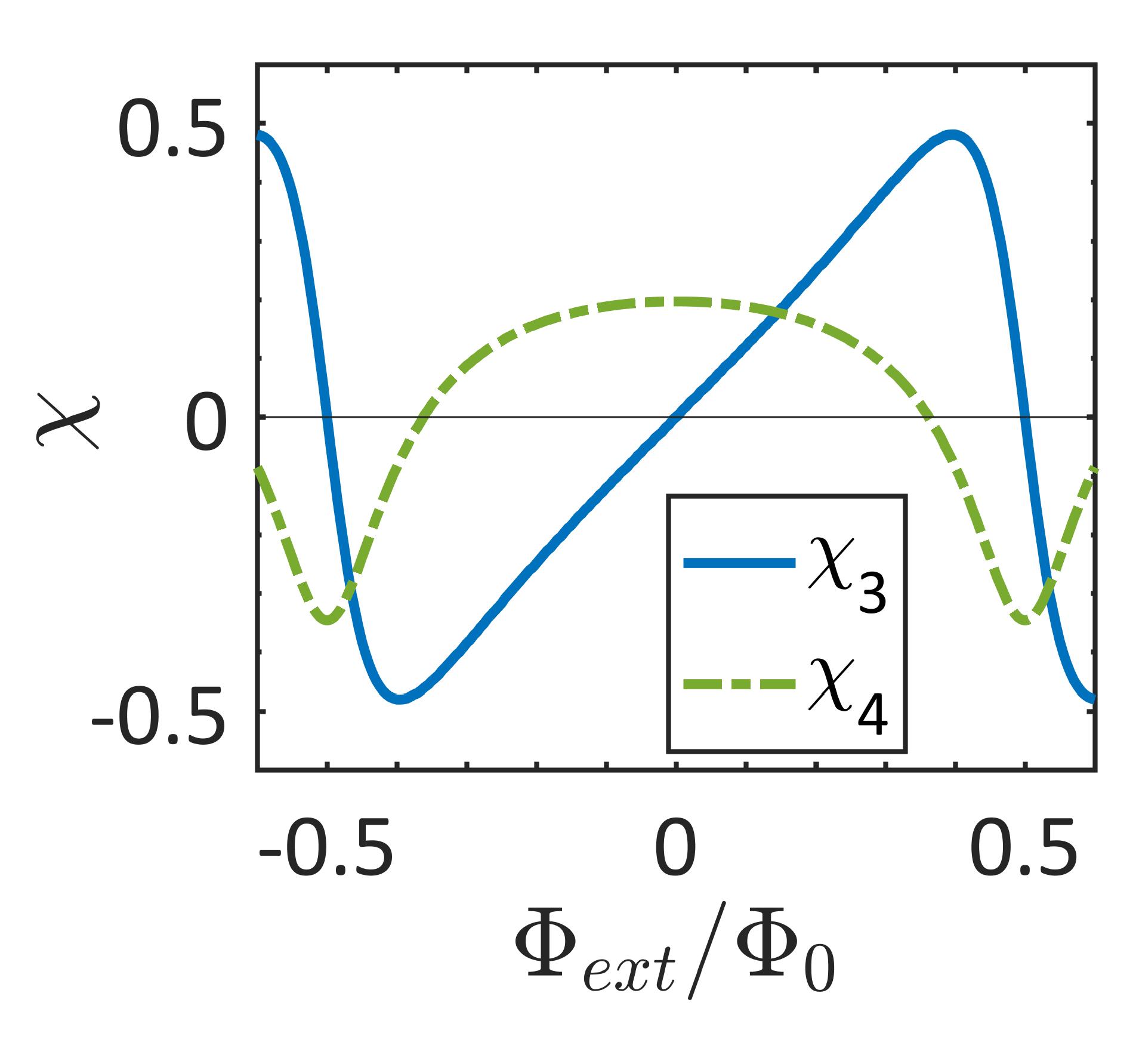}
         \caption{}
         \label{fig:C3C4}
     \end{subfigure}
    \caption{\small{Design of a stop-band by periodic loading in the TWPA chain. (a) Three cells made with SNAIL loops with inductances $L_1,\, L_2= (3/2)L_1$ and $C_1 = C_2 = C$ form a supercell. (b) The calculated dispersion diagram of the device consists of three propagation bands separated by two stop-bands. The pump frequency $\omega_p$ is chosen such that its second harmonic is within the stop-band. $\omega_c$ is the cutoff frequency and $k$ is the imaginary part of the complex propagation constant. (c) Calculated transmission of the TWPA ($S_{21}$) as a function of frequency and magnetic flux, $\Phi_{ext}$. The propagation bands, yellow, exhibit fringes resulting from the discrete structure of the TWPA; the stop-bands are shown in deep blue. $\Phi_0$ is the flux quantum. (d) Calculated nonlinearity coefficients $\chi_3$ and $\chi_4$, which define the strength of the 3WM and 4WM, respectively, as a function of the applied dc magnetic flux. }}
    \label{fig:design_SNAIL}
\end{figure*}

Our proposed solution to the problem of high harmonic generation is to create a stop-band for the second harmonic of the pump, $2\omega_p$, thus suppressing the generation and propagation of all pump harmonics. This is similar to a dispersion engineering approach for the 4WM-kinetic-inductance TWPAs, \cite{eom_2012, bockstiegel_2014, adamyan_2016} where a stop-band is used to suppress the third harmonic of the pump. A strong nonlinear dispersion in the vicinity of the stop-band would also prevent up-conversion of signal harmonics by the pump. Therefore, we anticipate a better amplifier performance as well as improved flexibility in  adjusting the pump frequency compared to other proposals. \cite{vissers_2016, Erickson_2017,malnou_2021} 

In our design, we use the method of periodic loading \cite{collin_2001} and modify the inductance of the chain to create a distributed filter, resulting in a stop-band at the desired frequency of $2\omega_p$. Such a filter is in the form of nonlinear inductor-capacitor ($LC$) cells, and besides its role in frequency elimination, it also contributes to the gain. Moreover, unlike resonator-based dispersion engineering, no extra element is added to the circuit, and the design is free from additional sources that can introduce reflection and loss. With this design, we demonstrate up to $20\, \text{dB}$ gain with a TWPA that contains only 440 $LC$ cells.

Illustrated in Fig.~\ref{fig:TWPA_UC} is the TWPA, where the nonlinearity results from the Josephson junctions, arranged in the form of a superconducting nonlinear asymmetric inductive element (SNAIL) loop, \cite{frattini_2017} to provide three-wave mixing. 
The equivalent SNAIL inductance, $L$, is periodically modulated such that the three consecutive cells form a \textit{supercell}, $L_1 - L_1 - L_2$, with $L_2= (3/2)L_1$ being the loaded feature. We choose the capacitance $C$ such that the supercell, as a whole, is impedance matched, while the variation of the TWPA impedance with magnetic flux is insignificant within the studied interval. In general, it is possible to place the loaded feature periodically after any number of $L_1$-cells and also to use a different capacitance, $C_2\neq C$. 

The SNAIL loops are configured with a single junction with Josephson energy $E_{J1}$ and three identical junctions, each with energy $E_{J2}$, $E_{J1}/E_{J2} = \alpha = 0.16$. For this configuration, the normalized SNAIL inductive energy is, \cite{frattini_2017} 
\begin{eqnarray}\label{SNAIL_energy}
{\frac{E_S (\tilde{\phi} )}{E_{J2}}} &=& -\alpha\cos(\tilde{\phi}+\tilde{\phi}_{min}) - 3\cos \Bigl(\frac{\phi_{ext} - (\tilde{\phi}+\tilde{\phi}_{min})}{3} \Bigr) \no
&\approx& c_0\tilde{\phi}_{min}+ c_2{\tilde{\phi}}^2 + c_3{\tilde{\phi}}^3 + c_4{\tilde{\phi}}^4 + \ldots\,,
\end{eqnarray}
where $\tilde{\phi}(t)$ is the phase difference across the SNAIL. The second line is an expansion of the energy around the minimum superconducting phase difference, $\tilde{\phi}_{min}$, defined by the biasing magnetic flux, $\phi_{ext} = 2\pi \Phi_{ext}/\Phi_0$. Equation \ref{SNAIL_energy} contains both third- and fourth-order nonlinear terms for the phase fluctuation, which are responsible for the 3WM and 4WM, respectively. 

The dispersion relation is obtained by calculating the transmission (\textit{ABCD}) matrix of the supercell, $\hat {\cal M}$, and solving the equation, $2\cosh(\gamma a) = {\rm Tr}(\hat {\cal M})$, where $\gamma$ is the complex propagation constant and $a$ is the length of the supercell. \cite{collin_2001} 
The dispersion relation is plotted in Fig.~\ref{fig:dispersion} for $\Phi_{ext} = 0$ ($k$ represents the imaginary part of $\gamma$). The plot consists of three propagation bands separated by two stop-bands. 
The variation of the stop-bands with magnetic flux is illustrated in Fig.~\ref{fig:S21}. The stop-bands shift downwards in frequency as the magnetic flux approaches $0.5\, \Phi_0$, because of the increase in the SNAIL inductance. 

To account for the nonlinear coupling of the waves, one needs to resort to a dynamical equation for the superconducting phase,
$\phi(x,t)$, representing the $n$-th node of the chain at $x= na$. For a homogeneous chain without loaded features, $L_2=L_1$, the equation has the form, \cite{nilsson_2022}
\begin{eqnarray}\label{EOM}
&&   {-CL_1} \ddot \phi - 4\sin^2\biggl({\frac{a\hat k}{2}}\phi\biggr) + {\frac{\chi_3}{2}} \hat{\cal D}_3[\phi] + 
{\frac{\chi_4}{3}}\hat{\cal D}_4[\phi] =0,  \no
&& \hat{\cal D}_{\mu}[\phi] = \left[\bigl(1-e^{-ia\hat k}\bigr)\phi \right]^{\mu-1} + \left[\bigl(e^{ia\hat k} - 1\bigr)\phi \right]^{\mu-1}, 
\end{eqnarray}
where $\hat k = -i\partial_x$, and the nonlinearity coefficients $\chi_3= -3c_3/c_2$ and $\chi_4= -6c_4/c_2$ quantify the strength of the 3WM and 4WM, respectively.
The dependence of these coefficients on the magnetic flux bias is shown in 
Fig.~\ref{fig:C3C4}. The 4WM coefficient $\chi_4$ vanishes at $\Phi_{ext}\approx 0.36\, \Phi_0$, while the $\chi_3$-coefficient at this point is close to its maximum value. For the loaded TWPA, the dynamical description is more complex; one has to resolve the nodal phases within the supercell, $n = 3m,\, 3m+1$, and $3m+2$, where $m$ is the supercell number. Then \Eq{EOM} splits into three coupled equations for the corresponding phases (cf.~[\onlinecite{nilsson_2022}]). Nevertheless, \Eq{EOM} remains useful for the loaded chain as a tool for qualitative description of a low-frequency region below the first stop-band. Furthermore, in the long-wavelength limit, $ka\ll1$ ($\omega \ll \omega_c$) and one can simplify \Eq{EOM} by Taylor expansion over the small parameter $ka$. 

\begin{figure}[t]
\centering
\captionsetup{justification=Justified}
 \includegraphics[width=0.50\textwidth]{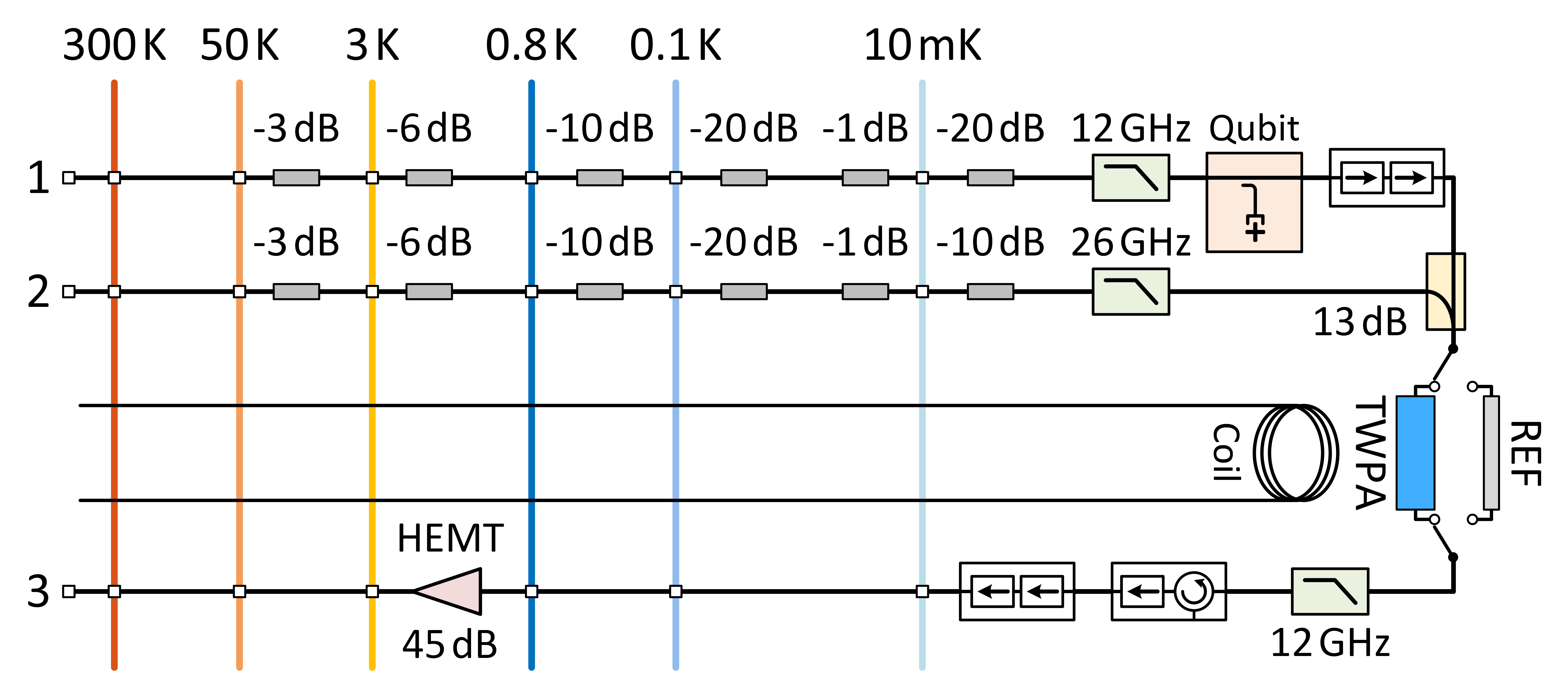}
         \caption{Schematic of the cryogenic measurement setup. The pump and the signal are sent through separate lines. The signal line contains a resonator with a qubit used for measuring the input power at the chip. The switchable bypass line (REF) across the TWPA is used as the reference.}
         \label{fig:wiring}
\end{figure}

We fabricated the device on a high resistivity silicon wafer ($\rho \geq 10\, \text{k}\Omega\, \text{cm}$), with evaporated aluminum as the superconducting material and aluminum/aluminum oxide/aluminum (Al/AlO$_x$/Al) Josephson junctions.  
We mounted a $5 \times 10\ \text{mm}^2$ TWPA chip, containing 440 cells or $\approx\, 147$ supercells, in a copper box and attached the assembly to the mixing chamber of a Bluefors dilution refrigerator to measure the device at temperatures around $10\, \text{mK}$. Figure~\ref{fig:wiring} shows the wiring diagram of the cryogenic measurement setup. 
To apply the dc magnetic flux, we placed a superconducting coil under the sample, and to calibrate the measurement, we used a $50\, \Omega$ SMA cable as the reference. The TWPA and the reference were connected to RF switches to share the input/output lines. 
The signal and the pump were fed through line 1 and line 2, respectively. They combined in a directional coupler before entering the TWPA.
This was made to protect a qubit, accessible via the signal line, from being perturbed by the strong pump. We used this qubit to calibrate the power, allowing us to measure the noise temperature, as discussed later. The qubit chip contains a fixed-frequency transmon qubit with the frequency $4.164\, \text{GHz}$, connected to a readout resonator with the frequency $6.0351\, \text{GHz}$. For gain measurements, we bypassed the qubit and instead combined the signal and the pump at room temperature and fed both through input line 2.

\begin{figure}[t]
         \centering
         \captionsetup{justification=Justified}
         \includegraphics[width=0.45\textwidth]{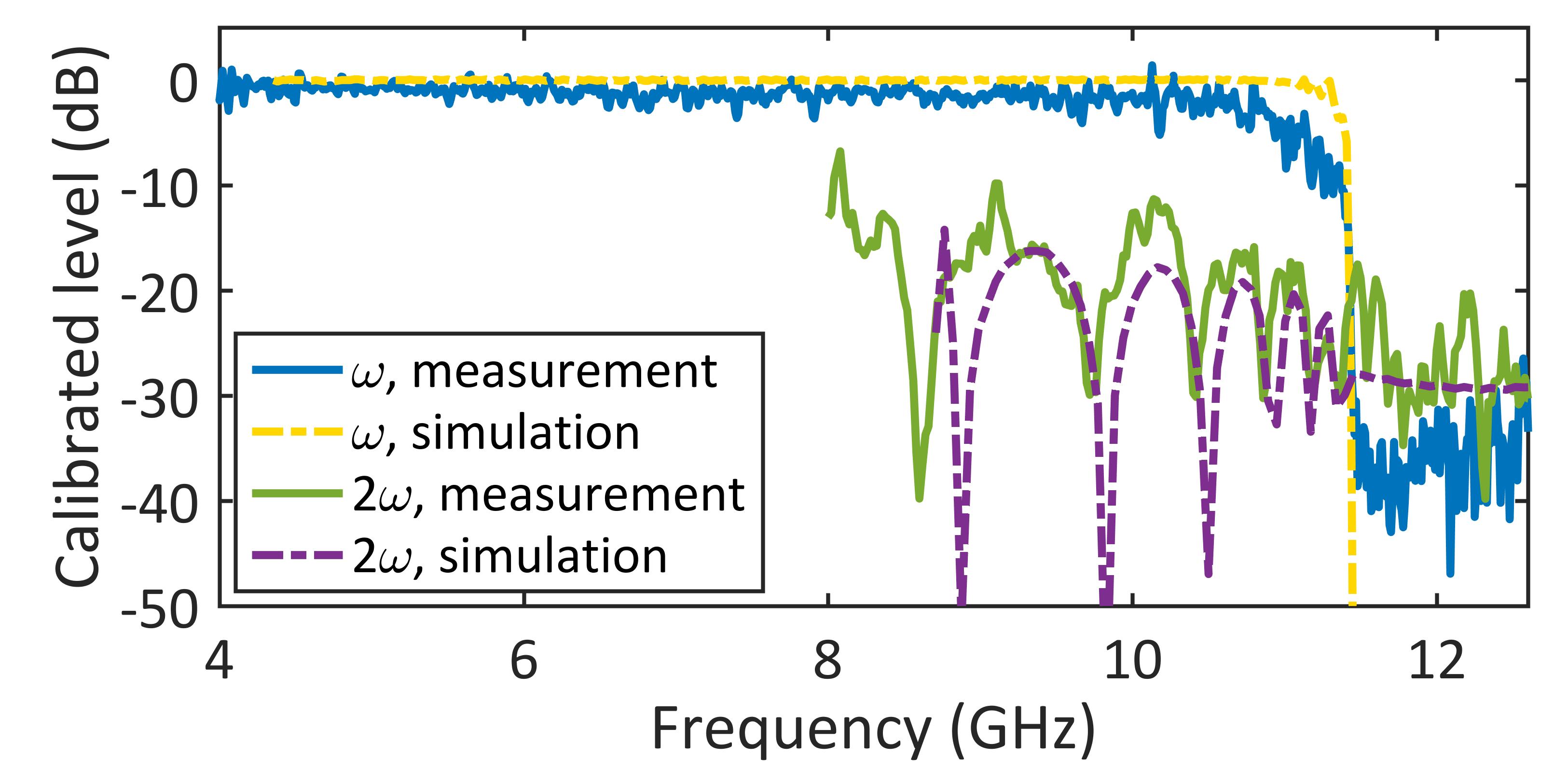}
  \caption{\small{ 
The frequency response of the TWPA to an input tone with power $-111\, \text{dBm}$ at $\Phi_{ext} \approx 0.38\, \Phi_0$. The response at the input frequency (blue) exhibits a stop-band at $f \gtrsim\, \text{11.5}\, \text{GHz}$, where the transmission drops by several orders of magnitude. The response at twice the input frequency (the second harmonic, green) also drops within the stop-band, but not as significantly as the main tone. The yellow and purple dashed-dotted lines indicate the simulation results described in the text.     
}}
\label{fig:f_2f_vs_f}
\end{figure}

To probe the propagation of an injected signal and its second harmonic under 3WM, we biased the TWPA at $\approx 0.38\, \Phi_0$ and sent in a tone at frequency $\omega$ and power $-111\, \text{dBm}$, and measured the output signals at $\omega$ and $2\omega$. The experimentally measured transmission of the injected tone and its second harmonic is shown in Fig.~\ref{fig:f_2f_vs_f}. The response is plotted after subtracting the background, taken at zero flux, where the stop-band is well above frequency $f=\omega/(2\pi) = 13\, \text{GHz}$, to get a response similar to that of a reference cable within the measurement frequency band.
The response of the tone itself (blue trace) reveals the stop-band starting at approximately $11.5\, \text{GHz}$. The amplitude of the second harmonic outside the stop-band (green trace) is smaller than the main tone by approximately $15\, \text{dB}$ or more and exhibits oscillations with a decreasing amplitude as the frequency increases. This is a characteristic behavior for a large magnitude of phase mismatch compared to the nonlinear coupling strength and can be qualitatively described based on the theory, Ref.~[\onlinecite{Armstrong_1962}]
\begin{eqnarray}\label{A2}
\left|{\frac{A_2(x)}{A_1(0)} }\right| \approx {\frac{1}{2}} \tanh \Biggl({\frac{4\epsilon_3}{ \Delta a}}\sin\biggl({\frac{\Delta x}{2}}\biggr)\Biggr) \,,
\end{eqnarray}
where $A_{1,2}$ are the amplitudes of the first and the second harmonic, $\Delta= k(2\omega) - 2k(\omega)$ is the phase mismatch, and $\epsilon_3 = |\chi_3A_1(0)|ka <\Delta a$ is the nonlinear coupling strength. 
The striking observation, however, is a small suppression of the second harmonic within the stop-band compared to its amplitude outside the stop-band. Our simulation quantifies this suppression by approximately a factor of three compared to the peaks, as illustrated by dashed-dotted lines in Fig.~\ref{fig:f_2f_vs_f}, while the main tone decreases by orders of magnitude. The simulation is performed using the harmonic-balance method implemented in the Advanced Design System simulator. \cite{ADS} 
This method accounts for the full nonlinear dynamics of the Josephson inductance in response to the applied ac current and magnetic flux and returns all inter-mixing products generated in the circuit. 

The difference in the behavior of the main tone and its second harmonic can be understood by noting that the second harmonic continuously interacts, even within the stop-band, with the main tone, which freely propagates at smaller frequencies well below the stop-band and continuously provides energy to the second harmonic.
The magnitude of the second harmonic within the stop-band can be estimated by considering the main tone as an effective source whose intensity is proportional to the nonlinear coupling strength $\epsilon_3$. This is comparable to the amplitude of the second harmonic propagating outside the stop-band, \Eq{A2}. 

%
\begin{figure}
    \centering
    \captionsetup[subfigure]{singlelinecheck=false,justification=raggedright,skip=0pt}%
    \begin{subfigure}{0.45\textwidth}
      \setbox0=\vbox{\caption{}\label{fig:gain_6GHz}}
      \sbox1{\includegraphics[width=0.95\textwidth]{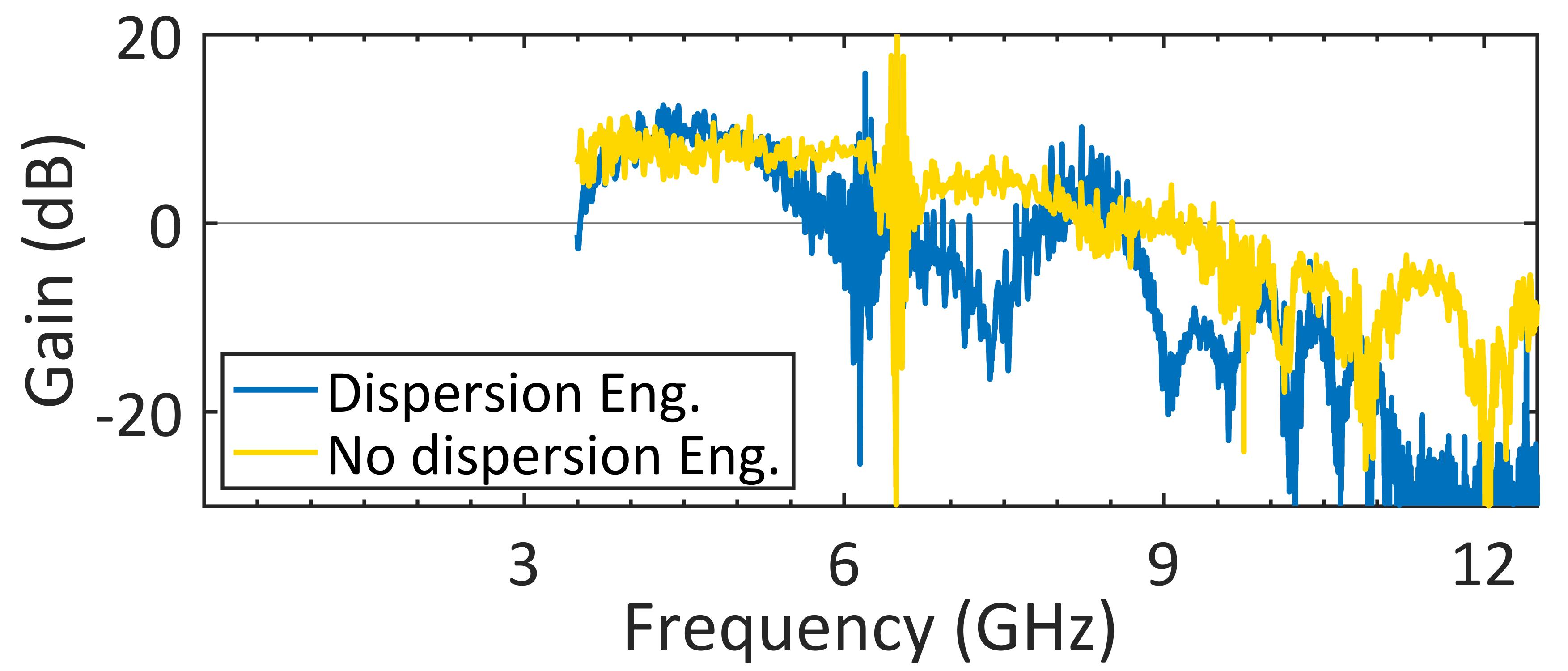}}
        \leavevmode\rlap{\usebox1}%
        \raisebox{\dimexpr \ht1-\ht5}{\usebox0}
    \end{subfigure}\hfil
    \begin{subfigure}{0.45\textwidth}
      \setbox0=\vbox{\caption{}\label{fig:modes_6GHz}}
      \sbox1{\includegraphics[width=0.95\textwidth]{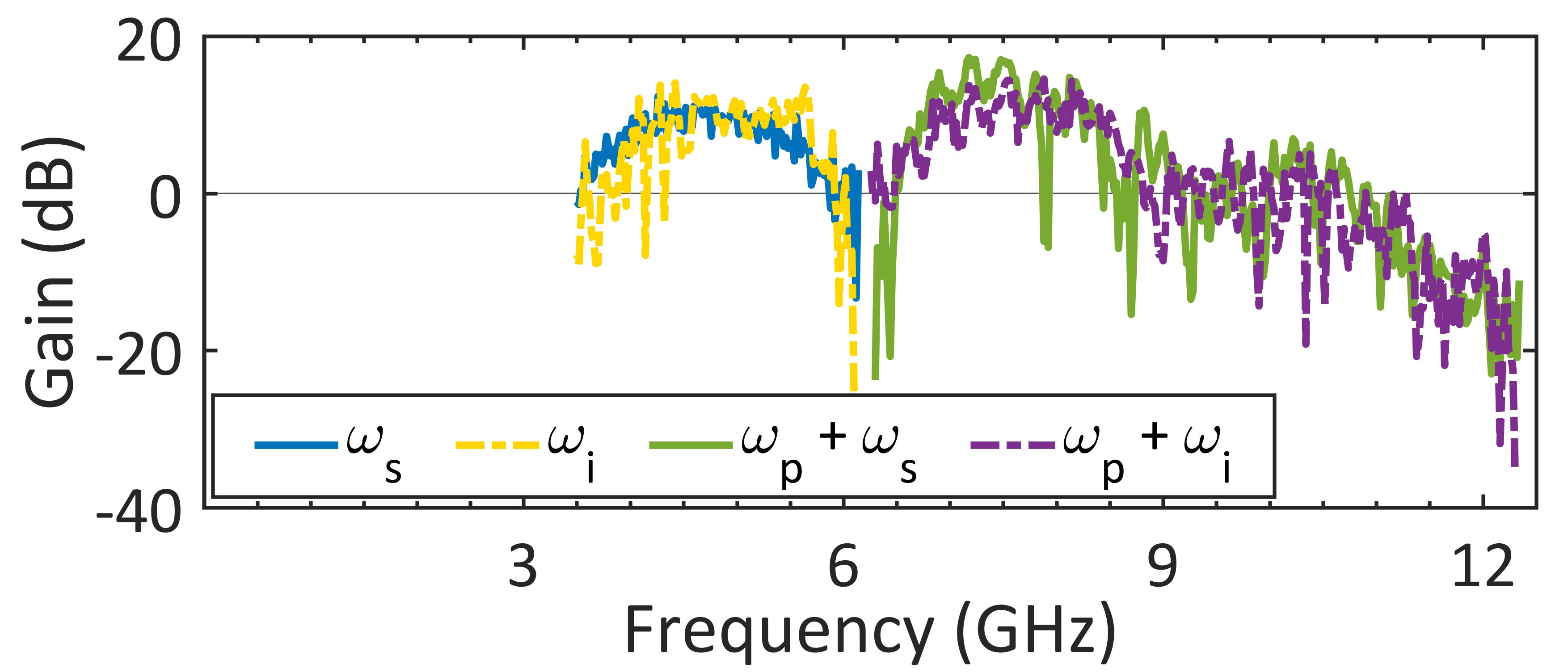}}
        \leavevmode\rlap{\usebox1}%
        \raisebox{\dimexpr \ht1-\ht5}{\usebox0}
    \end{subfigure}\hfil
    \begin{subfigure}{0.45\textwidth}
      \setbox0=\vbox{\caption{}\label{fig:upconversion_6GHz}}
      \sbox1{\includegraphics[width=0.95\textwidth]{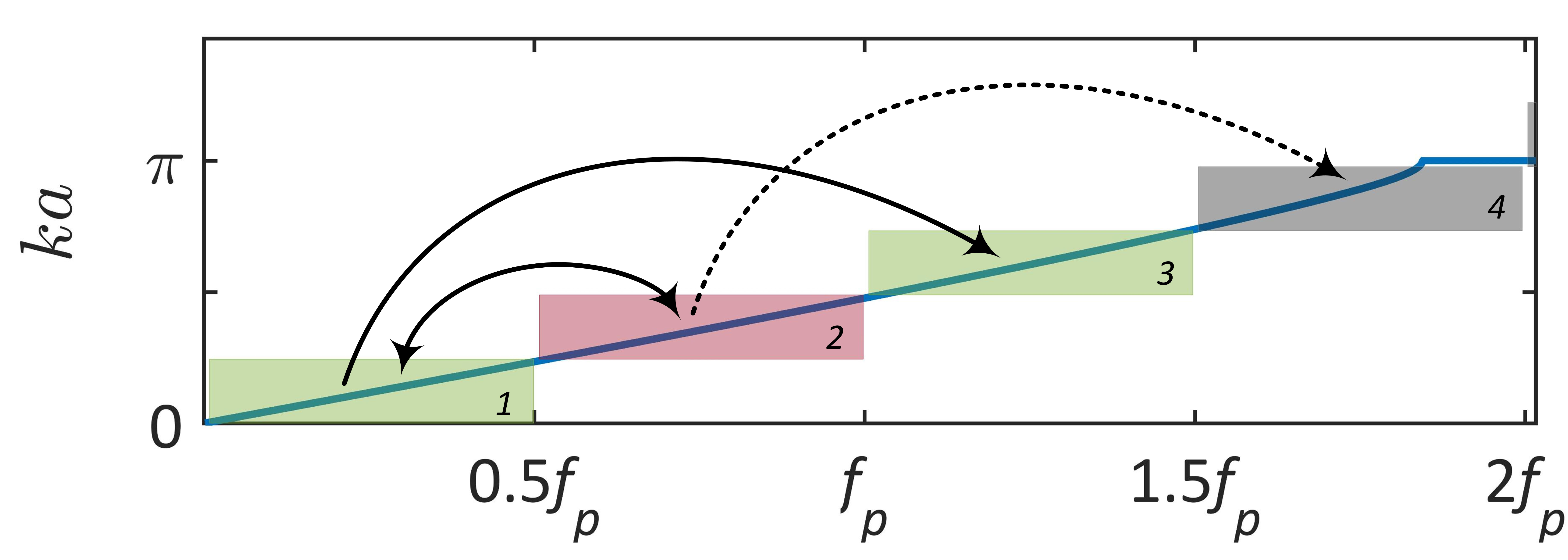}}
        \leavevmode\rlap{\usebox1}%
        \raisebox{\dimexpr \ht1-\ht5}{\usebox0}
    \end{subfigure}
    \captionsetup{justification=Justified}
    \caption{\small{Frequency dependence of the gain and harmonic generation under 3WM. $\Phi_{ext} \approx 0.38\, \Phi_0$,  $f_p=6.2\, \text{GHz}$, pump power, $P_p=-91.4\, \text{dBm}$.  
(a) Gain of the loaded TWPA, blue trace, in comparison with the gain of a TWPA without loading ($L_2 = L_1$), with the same number of cells (440), yellow trace. To account for the insertion loss, we compare the signal when the pump is on to that of the $50\, \Omega$ reference cable. 
(b) Frequency dependence of the output signal and 3WM idlers. The signal is injected within the frequency interval, $0.1$ - $6.2\, \text{GHz}$, while the output is detected within interval $3.5$ - $12.5\, \text{GHz}$. Blue trace refers to the detected signal, yellow - down-converted idler, green and purple - up-converted signal and idler, respectively. (c) Frequency diagram illustrating mode coupling: the modes within the lower green (1) and pink (2) frequency intervals are strongly coupled by down-conversion and are therefore amplified. The modes within the lower green (1) interval are efficiently up-converted to the green (3) region above the pump frequency, while the modes within the pink (2) region are not up-converted because of the stop-band and strong dephasing near the stop-band edge (gray region, 4).}}
    \label{fig:gain-upconversion}
\end{figure}

\begin{figure}[t]
\centering
\captionsetup{justification=Justified}
\includegraphics[width=0.425\textwidth]{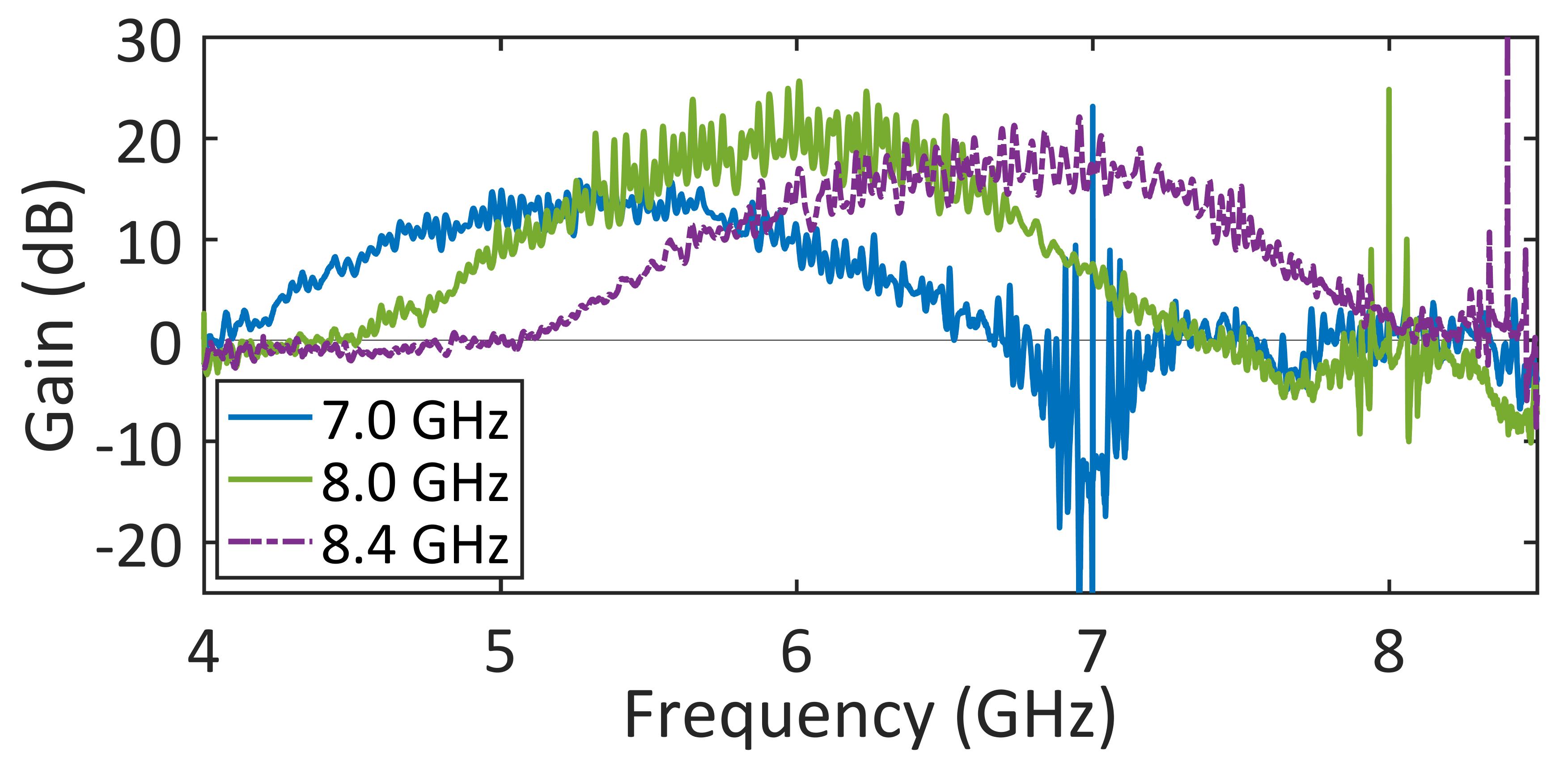}
\caption{\small{The gain of the TWPA at different pump frequencies. The flux bias is tuned such that the second harmonic of the pump is in the stop-band, and the pump power is also adjusted following the change of the critical current with flux. Blue: $f_p = 7.0\, \text{GHz}$, $\Phi_{ext}=0.34\, \Phi_0$, $P_p=-87.4\, \text{dBm}$;  
green: $f_p = 8.0\, \text{GHz}$, $\Phi_{ext}=0.22\, \Phi_0$, $P_p = -82.25\, \text{dBm}$, 
purple: $f_p = 8.4\, \text{GHz}$, $\Phi_{ext}=0.21\, \Phi_0$,  $P_p = -82.4\, \text{dBm}$.  
The gain is obtained with respect to the $50\, \Omega$ reference.}
}   
\label{fig:gain_multi_f}
\end{figure}
Despite the incomplete extinction of the second harmonic, the signal gain considerably changes, as our next observations show. We biased the device at $\Phi\approx0.38\, \Phi_0$, where 4WM is negligible, and applied a pump tone at $6.2\, \text{GHz}$ so that its second harmonic fell in the stop-band, and measured the signal gain at various frequencies. The result is shown in Fig.~\ref{fig:gain_6GHz} with the blue trace. The yellow trace represents the gain of a SNAIL-TWPA that contains no dispersion-engineering features. The latter is made of the same number of 440 SNAILs with similar configuration, and the measurements were done at the same operational point. The noticeable difference in the frequency dependence of the gain as well as some enhancement of the gain, indicate the effect of the suppression of high harmonics. To investigate the presence of the up-converted harmonics, we applied a signal within the interval $0.1$ - $6.2\, \text{GHz}$ and measured the response between $3.5$ - $12.5\, \text{GHz}$ (the available band of our equipment). In Fig.~\ref{fig:modes_6GHz} the blue trace indicates the measured response at the signal frequency, $\omega_s$, while the yellow trace depicts the (down-converted) idler, $\omega_i = \omega_p-\omega_s$, when the signal is applied below $2.7\, \text{GHz}$. The green and purple traces represent the up-conversion of the signal and idler, respectively, $\omega_p+\omega_{s,i}$.
The peak at frequencies between $6.2$ - $9\, \text{GHz}$ indicates efficient up-conversion from the quasi-linear dispersion region below $3\, \text{GHz}$, while the significantly weaker response at higher frequencies ($> 9\, \text{GHz}$) reveals the increased phase mismatch close to the stop-band. Therefore, we conclude that the shape of the gain in  Fig.~\ref{fig:gain_6GHz} is mainly formed by an interplay of the three modes, $\omega_s$, $\omega_i$, and $\omega_p+\omega_{s(i)}$, as shown in Fig.~\ref{fig:upconversion_6GHz}. 
The frequency shift of the gain peak from half of the pump frequency can be understood on the basis of the theory in Refs.~[\onlinecite{tien_1958}] and [\onlinecite{nilsson_2022}], where the maximum gain splits and shifts from $\omega_p/2$ due to strong dephasing. 

The shape of the gain traces remains qualitatively similar when the pump frequency changes. In Fig.~\ref{fig:gain_multi_f} we present the gain for three pump frequencies between $7$ and $8.4\, \text{GHz}$. For each plot, we kept the pump second harmonic within the stop-band, thus we changed the flux bias from $0.34\, \Phi_0$ down to $0.21\, \Phi_0$. We also ramped up the pump power taking advantage of the increased critical current. With the growing pump frequency, the gain increases, following the theory prediction. \cite{tien_1958, nilsson_2022} While the gain maximum is $\approx 15\, \text{dB}$ at the pump frequency of $7\, \text{GHz}$, it increases to about $20\, \text{dB}$ within the pump frequency interval $7.8$ - $8.4\, \text{GHz}$. The shape of the gain traces does not change with decreasing the magnetic flux bias, indicating that 3WM remains dominant within the range of applied magnetic bias, despite the increasing 4WM admixture. The data presented in Fig.~\ref{fig:gain_multi_f} is the main result of this work.\par

\begin{figure}[thb]
    \centering
    \captionsetup[subfigure]{singlelinecheck=false,justification=raggedright,skip=0pt}%
    \centering
    \begin{subfigure}{0.4\textwidth}
      \setbox0=\vbox{\caption{}\label{fig:dSNR_NoiseT}}
      \sbox1{\includegraphics[height=3.5cm]{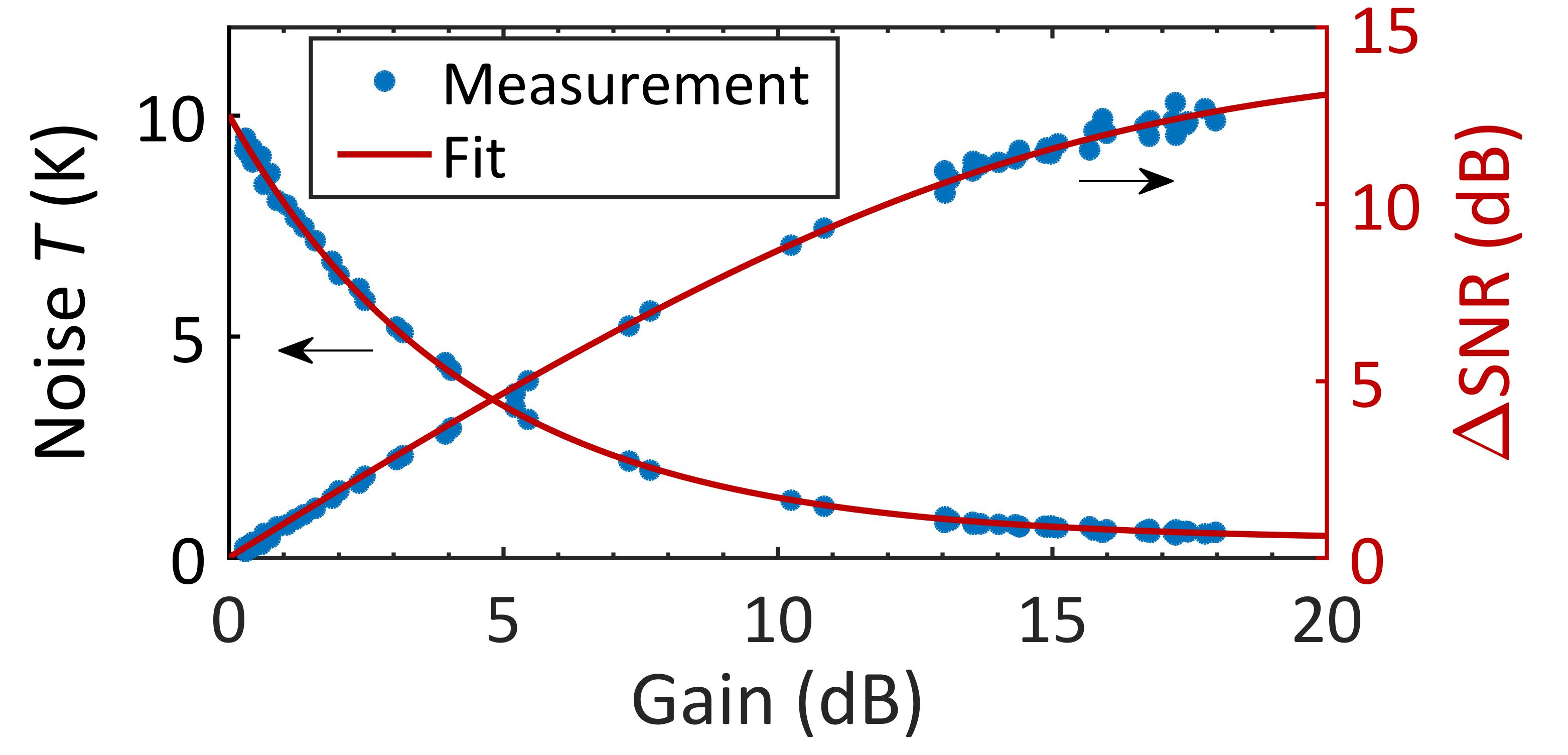}}
        \leavevmode\rlap{\usebox1}
        \raisebox{\dimexpr \ht1-\ht5}{\usebox0}
    \end{subfigure}\hfil
    \begin{subfigure}{0.4\textwidth}
      \setbox0=\vbox{\caption{}\label{fig:A_twpa}}
      \sbox1{\includegraphics[height=3.5cm]{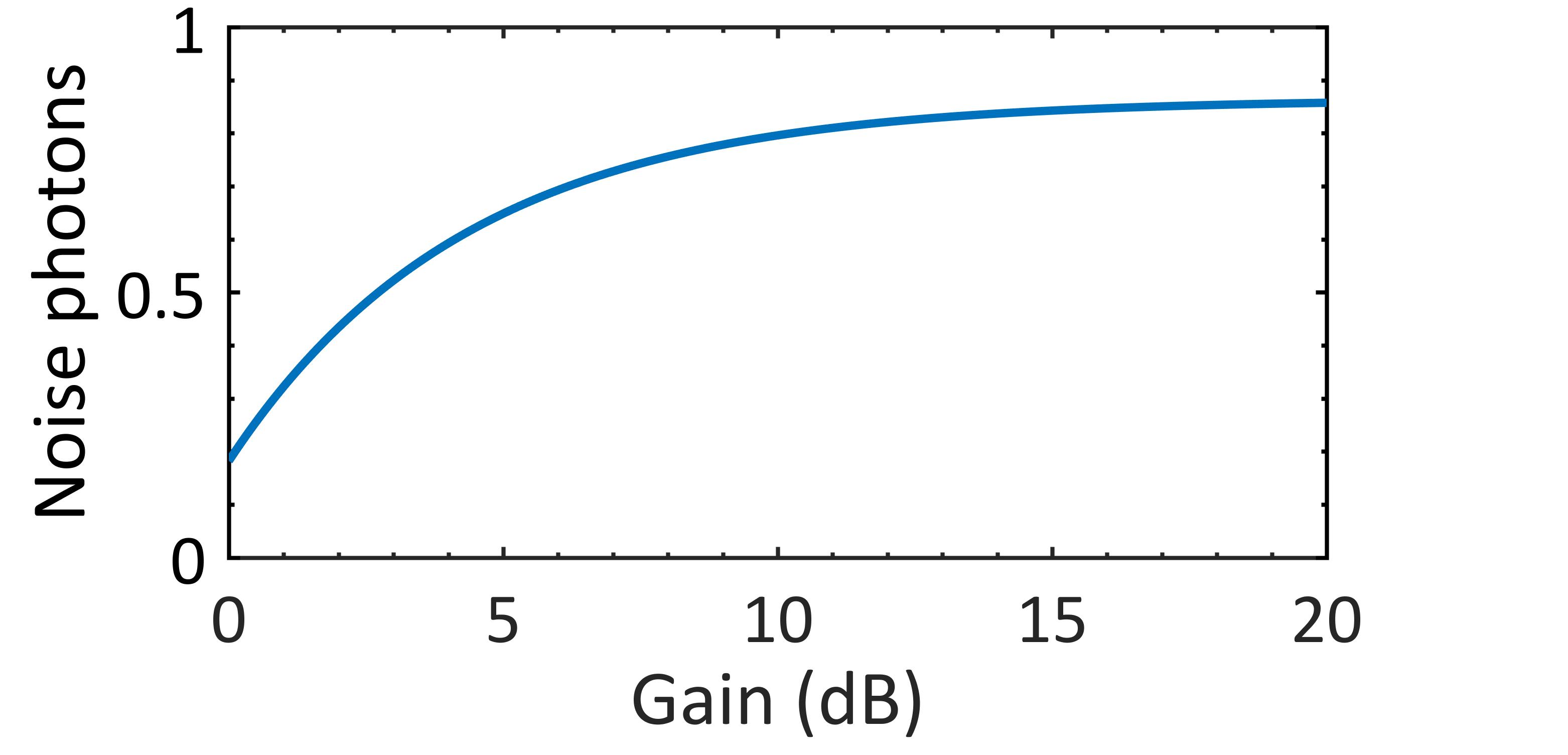}
         \begin{picture}(0,0)
         \put(-145,35){\includegraphics[height=1.2cm]{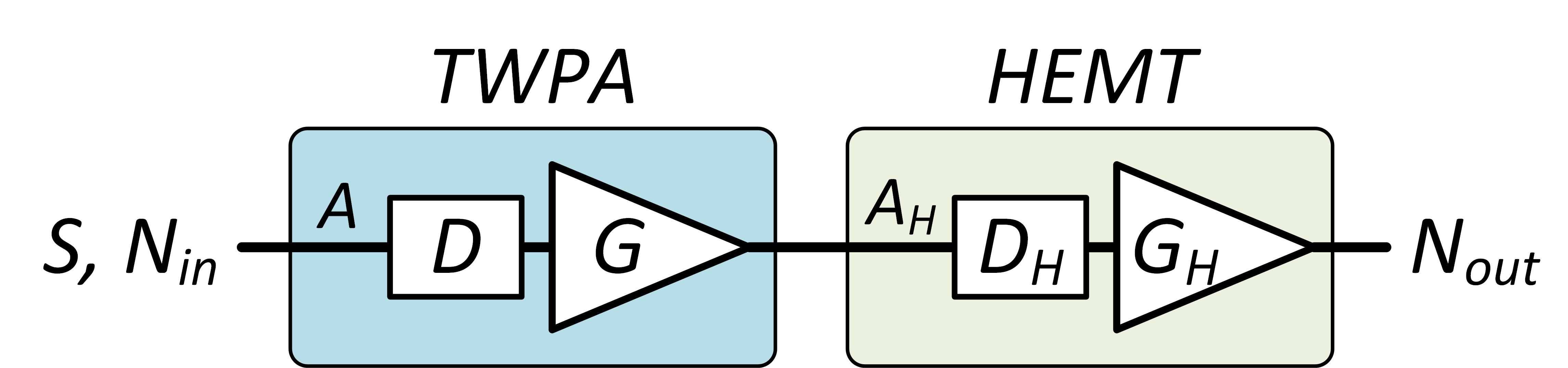}}
         \end{picture}
      }
        \leavevmode\rlap{\usebox1}
        \raisebox{\dimexpr \ht1-\ht5}{\usebox0}
    \end{subfigure}
    \captionsetup{justification=Justified}
    \caption{\small{Noise performance of the TWPA. (a) Noise temperature (left axis) and the signal-to-noise ratio improvement (right axis) of the system as a function of the TWPA gain. $f_p = 8.0\, \text{GHz}$, and the measurement is carried at $6.034\, \text{GHz}$. The fit is shown in solid red. When the TWPA is bypassed, the noise temperature is $6.5\, \text{K}$ (not shown in the figure). (b) The added noise photons of the TWPA, as a function of gain. Inset shows modeling the TWPA as an ideal amplifier with gain $G$, connected to a lossy element with damping $D$ in the front, used to make the fit in (a). At $G =0\, \text{dB}$, the non-zero number of noise photons is due to damping $D$, as $A = (1-D)/(2D)$. }}
       \label{fig:NnoiseT}
\end{figure}

%
We conclude with a discussion about the noise performance of the TWPA.
Figure~\ref{fig:dSNR_NoiseT} presents the noise temperature as a function of the gain, where the data points correspond to four close values of the magnetic flux bias $\Phi_{ext} \approx 0.2\, \Phi_0$. 
The input signal was calibrated by measuring the number of photons in the readout resonator ($f_r = 6.0351\, \text{GHz}$) of the qubit chip, see Fig.~\ref{fig:wiring}, via the ac Stark shift measurement. \cite{bruno_2015} In Fig.~\ref{fig:dSNR_NoiseT} we also depict the improvement of the signal-to-noise ratio ($\Delta$SNR) quantified with the ratio of the SNRs for the pump on and off.
The noise temperature of the system is about $10\, \text{K}$ when the TWPA is connected but not pumped, and drops to less than $0.6\, \text{K}$ when the TWPA is turned on. 
To analyze the data, we adopt a model for the TWPA as an ideal amplifier with gain $G$, connected to a lossy element with damping $D$ on the input, \cite{belove_1986_book} as shown in the inset of Fig.~\ref{fig:A_twpa}. $D$ includes the total damping between the qubit and the input of the TWPA as well. The added number of noise photons referred to the TWPA input is then $A = (1-D)/(2D) +(G-1)/(2GD)$, where the first term results from the damping, and the second term is due to the gain, assuming the standard quantum limit. \cite{Caves1982}
The total system noise of the TWPA connected in series to the HEMT amplifier is    
\begin{eqnarray}\label{A}
N_{tot} = N_{in} + \frac{G(2-D)-1}{2GD} + \frac{A_H}{GD}.
\end{eqnarray}
Here $N_{in}$ and $A_H$ are the input noise and the added noise by the HEMT amplifier, respectively, in units of noise photons. The second term is the noise added by the TWPA. The signal-to-noise ratio improvement is then
\begin{eqnarray}\label{SNRI}
\Delta \text{SNR}(G) = {\frac{N_{in} + A(G=1) + {\frac{A_H}{D}}}{N_{in} + A(G) + {\frac{A_H}{GD}}}}.
\end{eqnarray}
We use Eqs. (\ref{A}) and (\ref{SNRI}) to fit the data in Fig.~\ref{fig:dSNR_NoiseT}, shown with solid lines in the plots. Assuming vacuum input noise, $N_{in}=0.5$, we extract the parameters $D=0.73$ and $A_H = 24.85$, and compute the added noise of the TWPA, presented in Fig.~\ref{fig:A_twpa}. The saturated value of the added noise at large pump power is found, $A_{\infty}=0.86$ photons. Thus the added noise by TWPA is only 0.36 photons above the standard quantum limit of $A_{SQL}=0.5$. \cite{Caves1982} \par
In summary, we observed a significant boost in the gain of a Josephson junction-based TWPA in the three-wave-mixing regime. The gain enhancement is achieved by engineering a stop-band in the frequency spectrum of the TWPA by introducing periodic loading, implemented via alternating the SNAILs inductance in the unit cells. Placing the stop-band at twice the pump frequency allows for suppressing the generation of the higher harmonics of the pump, the signal and the idler. With a small number of 440 SNAILs (147 supercells), the device shows a maximum gain of $20\, \text{dB}$, about $10\, \text{dB}$ larger than the maximum gain of the TWPA with no dispersion engineering, a 3-dB bandwidth of $\approx 1\, \text{GHz}$ between $4$ and $8\, \text{GHz}$, and an added noise below one photon. The width of the stop-band provides the flexibility of tuning the pump frequency over a range of $1.4\, \text{GHz}$, while the maximum gain remains above $15\, \text{dB}$.

This research was funded by the Knut and Alice Wallenberg Foundation through the Wallenberg Center for Quantum Technology (WACQT) and by the EU Flagship on Quantum Technology (OpenSuperQ project). 
The authors are grateful to Christian Fager for valuable advice on device modeling. The authors acknowledge the use of the Nanofabrication Laboratory (NFL) at Chalmers University of Technology. The authors are thankful for the support from Marco Scigliuzzo, Andreas Bengtsson, Yong Lu, Liangyu Chen, Christopher Warren, David Niepce, Eleftherios Moschandreou, Philip Krantz and Lars Jönsson. \\

The data that support the findings of this study are available from the corresponding author upon reasonable request.\\

\section*{References}
\bibliography{references_APL.bib}
\end{document}